\documentclass[preprint,showpacs,preprintnumbers,amsmath,amssymb]{revtex4}

\usepackage{graphicx}
\usepackage{dcolumn}
\usepackage{bm}
\usepackage{subfigure}
\usepackage{theorem}
\newcommand{\mib}[1]{\mbox{\boldmath $#1$}}
\newcommand{\qed}{\hbox{\rule[-2pt]{3pt}{6pt}}}

\def\R{\mathbb{R}}
\def\N{\mathbb{N}}
\def\W{\mathbb{W}}
\def\C{\mathbb{C}}

\def\0{{\bf 0}}

\def\cG{{\cal G}}
\def\cS{{\cal S}}
\def\cN{{\cal N}}

\def\x{\mib{x}}
\def\y{\mib{y}}
\def\z{\mib{z}}
\def\k{\mib{k}}
\def\B{\mib{B}}
\def\K{\mib{K}}
\def\vnu{\mib{\nu}}
\def\vrho{\mib{\rho}}
\def\vlambda{\mib{\lambda}}
\def\vmu{\mib{\mu}}
\def\T{\mib{T}}
\def\Det{{\rm Det}}

\begin{document}

\preprint{}

\title{Noncolliding Brownian Motion with Drift 
and Time-Dependent Stieltjes-Wigert Determinantal Point Process}

\author{Yuta Takahashi}
\email{ytakahashi@phys.chuo-u.ac.jp}
\author{Makoto Katori}
\email{katori@phys.chuo-u.ac.jp}
\affiliation{%
Department of Physics,
Faculty of Science and Engineering,
Chuo University, 
Kasuga, Bunkyo-ku, Tokyo 112-8551, Japan 
}%

\date{28 September 2012}

\begin{abstract}
Using the determinantal formula of Biane, Bougerol, and O'Connell,
we give multitime joint probability densities 
to the noncolliding Brownian motion with drift,
where the number of particles is finite.
We study a special case such that the initial positions
of particles are equidistant with a period $a$
and the values of drift coefficients are well-ordered with a scale $\sigma$.
We show that, at each time $t >0$, the single-time probability
density of particle system is exactly transformed to
the biorthogonal Stieltjes-Wigert matrix model in the
Chern-Simons theory introduced by Dolivet and Tierz.
Here one-parameter extensions ($\theta$-extensions)
of the Stieltjes-Wigert polynomials,
which are themselves $q$-extensions of the Hermite polynomials,
play an essential role.
The two parameters $a$ and $\sigma$ of the process
combined with time $t$ are mapped to the
parameters $q$ and $\theta$ of the biorthogonal polynomials.
By the transformation of normalization factor
of our probability density, the partition function
of the Chern-Simons matrix model is readily obtained.
We study the determinantal structure of the matrix model
and prove that, at each time $t >0$,
the present noncolliding Brownian motion with drift
is a determinantal point process, in the sense that
any correlation function is given by a determinant
governed by a single integral kernel
called the correlation kernel.
Using the obtained correlation kernel,
we study time evolution of the noncolliding 
Brownian motion with drift.

\end{abstract}


\maketitle

\section{Introduction}
\label{chap:introduction}

Vicious walker models on lattices \cite{Fis84}
and their continuum versions \cite{Dys62}, 
which will be called noncolliding diffusion processes
\cite{KT07,KT_Sugaku_11},
have been extensively studied in connection with
the random matrix theory 
\cite{NF98,FNH99,Gra99,Baik00,Joh02,NF02,KT02,NKT03,KT04,KT10},
the Tracy-Widom distributions 
and extreme value statistics 
\cite{TW94,TW96,TW04,TW07,KIK08a,SMCR08,KIK08b,FMS11,Lie12},
the enumerative combinatorics and representation theory
\cite{GOV98,KGV00,ABE10,Fei12,ABE12},
the Riemann-Hilbert problem \cite{KMW09,DKZ11,KMW11}, 
the renormalization group theory \cite{CK03,GG10a,GG10b},
the growth models \cite{PS02,Joh03,IS05},
the quantum integrable systems \cite{Kat11,OCo12a,Kat12a},
and others.
Construction of noncolliding diffusion processes has been
based on the determinantal formula for nonintersecting paths
of Karlin-McGregor (KM) \cite{KM59}
and Lindstr\"om-Gessel-Viennot (LGV) \cite{Lin73,GV85}.
It can be regarded as a stochastic version of 
Slater's determinantal wave function for free fermions
in quantum mechanics 
and it means that indistinguishability of particles
({\it i.e.} invariance of statistics
under any exchange of paths at intersecting points 
on the spatio-temporal plane) should be assumed.
In general, if each walker (diffusion particle)
has different drift, then they are distinguishable
from each other and the KM-LGV determinantal formula
is not applicable.

If the values of drift coefficients of many particles 
are well-ordered, however, 
Biane, Bougerol and O'Connell \cite{BBO05}
proved that the following determinantal formula is valid.
Let $N=2,3, \dots$, 
and we consider an $N$-particle system of Brownian motions
in the one-dimensional real space $\R$ 
such that the $j$-th Brownian particle starts
at $x_j \in \R$ at time $t=0$ and
it has a constant drift $\nu_{j}$ per time,
$1 \leq j \leq N$.
In other words, we consider an $N$-dimensional
vector-valued diffusion process
\begin{equation}
\B^{\x, \vnu}(t)=\B^{\x}(t)+ \vnu t,
\quad t \geq 0,
\label{eqn:BM1}
\end{equation}
whose $j$-th component describes the
$j$-th Brownian motion with a drift coefficient $\nu_j$, 
\begin{eqnarray}
B^{x_j, \nu_j}_j(t) &=&
B^{x_j}_j(t)+\nu_j t
\nonumber\\
&=& x_j+B_j(t)+\nu_j t, \quad t \geq 0, \quad
1 \leq j \leq N,
\label{eqn:BM2}
\end{eqnarray}
where $B_j(t), t \geq 0, 1 \leq j \leq N$,
are independent one-dimensional standard Brownian motions
all started at the origin.
Let 
\begin{equation}
\W_N=\{\x =(x_1, x_2, \dots, x_N) \in \R^N :
x_1 < x_2 < \cdots < x_N \},
\label{eqn:Weyl}
\end{equation}
which is called the Weyl chamber of type A$_{N-1}$
in the representation theory \cite{FH91}.
Biane, Bougerol and O'Connell (BBO)
showed that \cite{BBO05}, if the values of the components
of initial configuration $\x=(x_1, x_2, \dots, x_N)$
and the drift vector $\vnu=(\nu_1, \nu_2, \dots, \nu_N)$
are in the same order, that is,
\begin{equation}
\x \in \W_N, \quad \vnu \in \W_N,
\label{eqn:order1}
\end{equation}
then the transition probability density of the drifted
Brownian motions (\ref{eqn:BM1}) 
{\it conditioned never to collide with each other}
is given by
\begin{equation}
p_N^{\vnu}(t, \y|\x)
=e^{-t|\vnu|^2/2}
\frac{\displaystyle{\det_{1 \leq j, k \leq N}[e^{\nu_j y_k}]}}
{\displaystyle{\det_{1 \leq j, k \leq N}[e^{\nu_j x_k}]}}
q_N(t, \y|\x), \quad \y \in \W_N, \quad t \geq 0,
\label{eqn:BBO1}
\end{equation}
where $|\vnu|^2=\sum_{j=1}^N \nu_j^2$ and
\begin{equation}
q_N(t,\y|\x)=
\det_{1 \leq j, k \leq N} \Big[p(t, y_j|x_k) \Big]
\label{eqn:KM1}
\end{equation}
with the transition probability density of $B(t)$, 
\begin{equation}
p(t,y|x)=\frac{1}{\sqrt{2 \pi t}} e^{-(y-x)^2/2t}, \quad t \geq 0.
\label{eqn:p0}
\end{equation}
(We give a brief review of the BBO argument \cite{BBO05}
in Appendix \ref{chap:appendix_BBO}.)

In the formula (\ref{eqn:BBO1}), we note that
even when some of $\nu_j$'s in 
$\vnu=(\nu_1, \dots, \nu_N)$ coincide, the ratio of determinants 
$\det_{1 \leq j,k \leq N}[e^{\nu_j y_k}]/
\det_{1 \leq j, k \leq N}[e^{\nu_j x_k}]$
can be interpreted using l'H\^opital's rule.
In particular, if we take the limit
$\nu_j \to 0, 1 \leq j \leq N$,
(\ref{eqn:BBO1}) is reduced to be
\begin{equation}
p_N(t, \y|\x)=
\frac{h_N(\y)}{h_N(\x)} q_N(t,\y|\x),
\quad \y \in \W_N, \quad t \geq 0
\label{eqn:KM2}
\end{equation}
with the Vandermonde determinant
\begin{equation}
h_N(\x)=\det_{1 \leq j, k \leq N} [x_j^{k-1}]
=\prod_{1 \leq j < k \leq N} (x_k-x_j).
\label{eqn:Vand1}
\end{equation}
It is considered as Doob's harmonic transform
($h$-transform) \cite{KT_Sugaku_11} of
the KM-LGV determinant $q_N(t, \y|\x)$ by $h_N$,
in which $h_N$ is harmonic in the sense that
$\Delta h_N(\x) \equiv \sum_{j=1}^N \partial^2 
h_N(\x)/\partial x_j^2=0$.
The above observation implies that
the BBO formula (\ref{eqn:BBO1}) 
is a generalization of the
$h$-transform of KM-LGV determinant (\ref{eqn:KM2}).

In (\ref{eqn:KM2}), we can take the limit
$x_j \to 0, 1 \leq j \leq N$,
which is denoted by $\x \to \0$,
and we obtain the probability density of
particle positions $\y=(y_1, y_2, \dots, y_N)$
of the noncolliding Brownian motions all started at
the origin,
\begin{equation}
p_N(t, \y|\0)=c_N(t)
(h_N(\y))^2 e^{-|\y|^2/2t},
\quad \y \in \W_N, \quad t \geq 0
\label{eqn:RM1}
\end{equation}
with $c_N(t)=t^{-N^2}/\{(2\pi)^{N/2}
\prod_{j=1}^N \Gamma(j)\}$.
The important fact is that (\ref{eqn:RM1})
can be identified with the probability density of 
eigenvalues of random matrices (ordered in $\W_N$) 
in the Gaussian unitary ensemble (GUE)
with variance $\sigma^2=t$ \cite{Meh04,For10}.
When $\x \not=\0$,
(\ref{eqn:KM2}) gives the probability
density of eigenvalues $\{y_j\}_{j=1}^N$
in the GOE-GUE two-matrix model 
studied in the high-energy physics \cite{Meh04},
in which the hermitian random matrix is
coupled with a real-symmetric random matrix 
with eigenvalues $\{x_j\}_{j=1}^N$
\cite{KT02,Kat12_pf}.
In this sense, the BBO formula (\ref{eqn:BBO1})
is expected to be related with some matrix-models
which are generalizations of two-matrix models.
(See \cite{BBO05,BBO09,OCo12b} for the relations
of (\ref{eqn:BBO1}) with the 
Duistermaat-Heckman measure
and the Littelmann path model in representation
theory.)

On the other hand, the BBO formula (\ref{eqn:BBO1})
is regarded as a simplified version of the 
transition probability density of the O'Connell process,
which is a stochastic version of
the quantum Toda lattice \cite{OCo12a}.
The reduction from the O'Connell process
to the noncolliding Brownian motion with drift
is called {\it a combinatorial limit} (or tropicalization) 
and the inverse procedure is called a
{\it geometric lifting} \cite{BBO09,OCo12b,Kat12b,Kat12c}.
(See Appendix \ref{chap:appendix_OCo}.)
In an earlier paper \cite{Kat12b},
we took the limit $\x \to \0$ with keeping
$\vnu \in \W_N$ and derived
a reciprocal time relation between the noncolliding
Brownian motion with drift $\vnu$ started at $\0$ 
and that without drift started at a configuration given by $\vnu$.
We would like to say that the BBO formula (\ref{eqn:BBO1})
is located in the high level in the random matrix theory
and it also gives an introduction to mathematical models
in higher levels \cite{OCo12b,BC11,Kat12c}.

In the present paper, we consider a special case
such that the drift coefficients and initial positions
are given by the following.
Let $n \in \N \equiv \{1,2, \dots,\}, N=2n-1$ and define
\begin{eqnarray}
\vrho &=&
\left( -\frac{N-1}{2}, -\frac{N-1}{2}+1, \dots,
-1, 0, 1 \dots, \frac{N-1}{2}-1, \frac{N-1}{2} \right)
\nonumber\\
&=& (-n+1, -n+2, \cdots, n-2, n-1).
\label{eqn:rho}
\end{eqnarray}
Then we put
\begin{eqnarray}
\label{eqn:equidistant}
&& x_j=a \rho_j= a(j-n)=a \left(j-\frac{N+1}{2} \right), 
\quad 1 \leq j \leq N,\\
\label{eqn:cond1}
&& \nu_j= \sigma \rho_j
= \sigma(j-n)=\sigma \left(j-\frac{N+1}{2} \right), 
\quad 1 \leq j \leq N,
\end{eqnarray}
with positive constants $a>0, \sigma >0$.
Although this setting seems to be very special,
we find that the obtained process can be
transformed to a matrix model
in Chern-Simons theory recently introduced
by Dolivet and Tierz \cite{DT07}.
Precisely speaking, the normalization factor
of the single-time probability density in our process
gives the inverse of partition function $Z$ of their matrix model.
Since here we study an interacting particle system
(the noncolliding Brownian motion with drift)
we can discuss not only the partition function
but also correlation functions.
Dolivet and Tierz shows that the ensemble of eigenvalues 
of their matrix model realizes 
the biorthogonal ensemble \cite{Mut95,Bor99}
associated with the one-parameter extension \cite{DT07} 
of the Stieltjes-Wigert polynomials \cite{Sze81,KS96}.
As an extension of their result, we will show in this paper
that this ensemble is a determinantal (fermion)
point process in the sense of \cite{Sos00,ST03}.
This implies that, at each time $t > 0$,
the noncolliding Brownian motion
started at the equidistant points (\ref{eqn:equidistant})
with the special drift (\ref{eqn:cond1}) is also
a determinantal point process, whose correlation kernel
is expressed by using the biorthogonal Stieltjes-Wigert 
polynomials.
We will report the time-evolution of the correlation 
functions of this noncolliding Brownian motion
with drift.

The paper is organized as follows.
The probability densities of the noncolliding Brownian motion
with drift and its transformations are given in Sec.\ref{chap:prob}.
The biorthogonal Stieltjes-Wigert polynomials are introduced
in Sec.\ref{chap:SW} and determinantal structure of the
ensembles are studied there.
In Sec.\ref{chap:evolution} the main result is stated and 
analytic and numerical
study of time-evolution of the present
noncolliding Brownian motion with drift using the correlation kernel is
reported. Concluding remarks will be given
in Sec.\ref{chap:conclusion}. Appendices are prepared for
the BBO argument, the O'Connell process, 
and the geometric Brownian motions, which are related
with the present process.

\section{Probability Densities of the Particle Systems}
\label{chap:prob}
\subsection{Multitime Joint Probability Density of
the Noncolliding Brownian Motion with Drift}
\label{section:multi}

In a previous paper \cite{Kat12b}, we considered an $N$-particle
system of noncolliding Brownian motion started at
$\x=(x_1, x_2, \cdots, x_N) \in \overline{\W}_N
\equiv \{\x \in \R^N: x_1 \leq x_2 \leq \cdots \leq x_N\}$
with a drift vector $\vnu=(\nu_1,\nu_2, \dots, \nu_N)$ 
satisfying $\vnu \in \overline{\W}_N$.
For any $M \in \N$
and an arbitrary $M$ sequence of times
$0 < t_1 < t_2 < \cdots < t_M < \infty$,
the multitime joint probability density of the process
is given by
\begin{eqnarray}
&& p_N^{\vnu}(t_1, \x^{(1)}; \dots; t_M, \x^{(M)}|\x)
\nonumber\\
&& = e^{-t_M|\vnu|^2/2}
\det_{1 \leq j, k \leq N} \left[e^{\nu_j x^{(M)}_k} \right]
\prod_{m=1}^{M-1} q_N(t_{m+1}-t_m, \x^{(m+1)}|\x^{(m)})
\frac{q_N(t_1, \x^{(1)}|\x)}
{\displaystyle{\det_{1 \leq j, k \leq N}
[e^{\nu_j x_k}]}},
\label{eqn:mpA1}
\end{eqnarray}
where particle configurations at each time $t_m$ is denoted by
$\x^{(m)}=(x^{(m)}_1, \dots, x^{(m)}_N) \in \W_N,
1 \leq m \leq M$. 
Here $q_N$ is the KM-LGV determinant
given by (\ref{eqn:KM1}). 
From this general formula, we can obtain the
following multitime joint probability density
for the special case (\ref{eqn:equidistant}) and 
(\ref{eqn:cond1}).

\vskip 0.5cm
\noindent{\bf Proposition 1} \quad
Consider the noncolliding Brownian motion
with $N$ particles
started at the equidistant points (\ref{eqn:equidistant})
at time $t=0$ with the drift having 
the coefficients (\ref{eqn:cond1}).
For an arbitrary $M \in \N$
and an arbitrary sequence of times
$0 < t_1 < t_2 < \cdots < t_M < \infty$,
the multitime joint probability density
of the process is given by 
\begin{eqnarray}
&& \widehat{p}_N(t_1, \x^{(1)}; \dots; t_M, \x^{(M)})
= c_N(a, \sigma, t_1, t_M) 
\prod_{\ell=1}^N (e^{\sigma x^{(M)}_{\ell}})^{-(N-1)/2}
\nonumber\\
&& \quad \quad \times
\prod_{1 \leq j < k \leq N}
(e^{\sigma x^{(M)}_k}-e^{\sigma x^{(M)}_j})
\prod_{m=1}^{M-1} q_N(t_{m+1}-t_{m}, \x^{(m+1)}|\x^{(m)})
\nonumber\\
&& \quad \quad \times
\prod_{j=1}^{N} p(t_1, x^{(1)}_j|0)
\prod_{\ell=1}^N (e^{a x^{(1)}_{\ell}/t_1})^{-(N-1)/2}
\prod_{1 \leq j< k \leq N}
(e^{a x^{(1)}_k/t_1}-e^{a x^{(1)}_j/t_1})
\label{eqn:mpB1}
\end{eqnarray}
with
\begin{equation}
c_N(a, \sigma, t_1, t_M)
=\frac{1}{\prod_{n=1}^{N-1} (e^{n a \sigma}-1)^{N-n}}
\exp \left\{ -\frac{1}{24} N(N^2-1) 
\left( \sigma^2 t_M-2 a \sigma + \frac{a^2}{t_1} \right) \right\}.
\label{eqn:cN1}
\end{equation}
It is also written as
\begin{eqnarray}
&& \widehat{p}_N(t_1, \x^{(1)}; \dots; t_M, \x^{(M)})
\nonumber\\
&& = c_N(a, \sigma, t_1, t_M) 
\prod_{1 \leq j < k \leq N}
\left[ 2 \sinh \frac{\sigma (x^{(M)}_k-x^{(M)}_j)}{2} \right]
\prod_{m=1}^{M-1} q_N(t_{m+1}-t_{m}, \x^{(m+1)}|\x^{(m)})
\nonumber\\
&& \qquad \qquad \times
\prod_{j=1}^{N} p(t_1, x^{(1)}_j|0)
\prod_{1 \leq j < k \leq N}
\left[ 2 \sinh \frac{a (x^{(1)}_k-x^{(1)}_j)}{2 t_1} \right].
\label{eqn:mpB1b}
\end{eqnarray}
\vskip 0.5cm
\noindent{\it Proof} \,
If we set the drift vector as (\ref{eqn:cond1})
with a positive constant $\sigma >0$, we obtain
$$
\det_{1 \leq j, k \leq N} [e^{\nu_j x_k}]
=\det_{1 \leq j, k \leq N} 
\Big[ (e^{\sigma x_k})^{j-1-(N-1)/2} \Big]
=\prod_{\ell=1}^N (e^{\sigma x_{\ell}})^{-(N-1)/2}
\prod_{1 \leq j < k \leq N} 
(e^{\sigma x_k}-e^{\sigma x_j}).
$$
We find that the above is equal to
$$
\prod_{1 \leq j < k \leq N}
\left[ 2 \sinh \frac{\sigma(x_k-x_j)}{2} \right].
$$
Note that 
$$
|\vnu|^2= \sigma^2 |\vrho|^2
= \sigma^2 \sum_{j=1}^{N} \left( j-\frac{N+1}{2} \right)^2
= \frac{1}{12} N(N^2-1) \sigma^2.
$$
Then the multitime joint probability density of the
process
with drift coefficients (\ref{eqn:cond1}) is given by
\begin{eqnarray}
&& p_N^{\sigma \vrho}
(t_1, \xi^{(1)}; \dots; t_M, \xi^{(M)}|\x)
\nonumber\\
&& \quad = e^{-N(N^2-1) \sigma^2 t_M/24}
\prod_{\ell=1}^N (e^{\sigma x^{(M)}_{\ell}})^{-(N-1)/2}
\prod_{1 \leq j < k \leq N}
(e^{\sigma x^{(M)}_k}-e^{\sigma x^{(M)}_j})
\nonumber\\
&& \qquad \times
\prod_{m=1}^{M-1} q_N(t_{m+1}-t_m, \x^{(m+1)}|\x^{(m)})
\frac{q_N(t_1, \x^{(1)}|\x)}
{\displaystyle{
\prod_{\ell=1}^{N} (e^{\sigma x_{\ell}})^{-(N-1)/2}
\prod_{1 \leq j < k \leq N}
(e^{\sigma x_k}-e^{\sigma x_j})}}
\nonumber\\
&& \quad = e^{-N(N^2-1) \sigma^2 t_M/24}
\prod_{1 \leq j < k \leq N}
\left[ 2 \sinh \frac{\sigma(x^{(M)}_k-x^{(M)}_j)}{2} \right]
\nonumber\\
&& \qquad \times 
\prod_{m=1}^{M-1} q_N(t_{m+1}-t_m, \x^{(m+1)}|\x^{(m)})
\frac{q_N(t_1, \x^{(1)}|\x)}
{\displaystyle{
\prod_{1 \leq j < k \leq N}
\left[ 2 \sinh \frac{\sigma(x_k-x_j)}{2} \right]
}}.
\label{eqn:ptilde1}
\end{eqnarray}

Then we assume that the initial configuration is given by
the equidistant points (\ref{eqn:equidistant})
with $a >0$.
We can see that in this setting
\begin{eqnarray}
&& q_N(t_1, \x^{(1)}|a \vrho)
\nonumber\\
&& \quad =e^{-N(N^2-1) a^2/24 t_1}
\frac{e^{-|\x^{(1)}|^2/2t_1}}{(2 \pi t_1)^{N/2}}
\det_{1 \leq j, k \leq N} 
\Big[ (e^{a x^{(1)}_k/t_1})^{j-1-(N-1)/2} \Big]
\nonumber\\
&& \quad = e^{-N(N^2-1) a^2/24 t_1}
\prod_{j=1}^{N} p(t_1, x^{(1)}_j|0)
\prod_{\ell=1}^N (e^{a x^{(1)}_{\ell}/t_1})^{-(N-1)/2}
\prod_{1 \leq j < k \leq N}
(e^{a x^{(1)}_k/t_1}-e^{a x^{(1)}_j/t_1})
\nonumber\\
&& \quad = e^{-N(N^2-1) a^2/24 t_1}
\prod_{j=1}^{N} p(t_1, x^{(1)}_j|0)
\prod_{1 \leq j < k \leq N}
\left[ 2 \sinh \frac{a (x^{(1)}_k-x^{(1)}_j)}{2 t_1} \right].
\nonumber
\end{eqnarray}
Moreover, we can show
$$
\prod_{\ell=1}^{N} (e^{\sigma a(\ell-(N+1)/2)})^{-(N-1)/2}=1,
$$
and
$$
\prod_{1 \leq j < k \leq N}
\{e^{\sigma a(k-(N+1)/2)}-e^{\sigma a(j-(N+1)/2)}\}
=e^{-N(N^2-1) a \sigma/12}
\prod_{n=1}^{N-1}(e^{n a \sigma}-1)^{N-n}.
$$
Therefore, we obtain the expressions (\ref{eqn:mpB1}) and
(\ref{eqn:mpB1b}) with (\ref{eqn:cN1}) for 
$\widehat{p}_N(\cdots) \equiv p_N^{\sigma \vrho}(\cdots|a \vrho)$. 
\qed

\subsection{Transformation to the System 
Associated with Geometric Brownian Motions}

Let $\R_+ \equiv (0, \infty)$. 
We change the variables as
$x^{(m)}_j \in \R \to y^{(m)}_j \in \R_+$ by
\begin{equation}
e^{\sigma x^{(m)}_j}=y^{(m)}_j \quad
\Longleftrightarrow \quad
x^{(m)}_j=\frac{1}{\sigma} \ln y^{(m)}_j, \quad
1 \leq m \leq M, \quad 1 \leq j \leq N, 
\label{eqn:change1}
\end{equation}
and put
\begin{equation}
\widehat{p}_N(t_1, \x^{(1)}; \dots ; t_M, \x^{(M)})
\prod_{m=1}^M d \x^{(m)}
= \widetilde{p}_N(t_1, \y^{(1)}; \dots; t_M, \y^{(M)})
\prod_{m=1}^M d \y^{(m)}
\label{eqn:change2}
\end{equation}
with $d \x^{(m)}=\prod_{j=1}^N dx_j^{(m)}$,
$d \y^{(m)}=\prod_{j=1}^N dy_j^{(m)}$, $1 \leq m \leq M$.
Then we obtain
\begin{eqnarray}
&& \widetilde{p}_N(t_1, \y^{(1)}; \dots; t_M, \y^{(M)})
= c_N(a, \sigma, t_1, t_M)
\prod_{\ell=1}^N (y^{(M)}_{\ell})^{-(N-1)/2}
\nonumber\\
&& \qquad \times
\prod_{1 \leq j<k \leq N}(y^{(M)}_k-y^{(M)}_j)
\prod_{m=1}^{M-1} q_N^{\rm geo}(t_{m+1}-t_m, \y^{(m+1)}|\y^{(m)})
\nonumber\\
&& \qquad \times
\prod_{j=1}^N p^{\rm geo}(t_1, y^{(1)}_j|1)
\prod_{\ell=1}^N (y^{(1)}_{\ell})^{-(N-1) \theta(t_1) /2}
\prod_{1 \leq j < k \leq N} 
\Big[(y^{(1)}_k)^{\theta(t_1)}-(y^{(1)}_j)^{\theta(t_1)} \Big]
\label{eqn:mpC1}
\end{eqnarray}
with
\begin{equation}
p^{\rm geo}(t, y|x)
=\frac{1}{y \sigma \sqrt{2 \pi t}}
\exp \left\{ - \frac{(\ln(y/x))^2}{2 \sigma^2 t} \right\}
\label{eqn:plog1}
\end{equation}
and
\begin{equation}
q^{\rm geo}_N(t, \y|\x)
=\det_{1 \leq j, k \leq N}
\Big[ p^{\rm geo}(t, y_j|x_k) \Big],
\label{eqn:qNlog1}
\end{equation}
where
\begin{equation}
\theta(t)=\frac{a}{\sigma t}.
\label{eqn:theta1}
\end{equation}
If we see (\ref{eqn:plog1}) as a function of $y$,
it is considered as the probability density
of the {\it log-normal distribution} with parameters
$\sigma \sqrt{t}$ and $\log x$.
As explained in Appendix \ref{chap:appendix_GBM}, 
it gives the transition probability
density from $x$ to $y$ in time duration $t > 0$
of the {\it geometric Brownian motion} $X(t), t \geq 0$,
which is given by an exponential of
one-dimensional standard Brownian motion $B(t)$; 
\begin{equation}
X(t)=x \exp (\sigma B(t)), \quad t \geq 0
\label{eqn:geoBM1}
\end{equation}
with the initial value $x=X(0)$
and with the parameter $\sigma >0$ 
called the percentage volatility \cite{BS02}.

It should be noted that, though (\ref{eqn:qNlog1})
is the KM-LGV determinant of
$p^{\rm geo}$'s, 
(\ref{eqn:mpC1}) is different from the
transition probability density of
`noncolliding geometric Brownian motion'. 
(See Appendix \ref{chap:appendix_noncGBM} for detail.)

\subsection{Single-Time Probability Density and Transformation
to Biorthogonal Ensemble}

From now on we consider only a single-time distribution. 
By setting $M=1$ and simplifying the notations as
$t=t_1, \y=\y^{(1)}$, 
(\ref{eqn:mpC1}) gives
\begin{eqnarray}
\widetilde{p}_N(t, \y) &=& c_N(a, \sigma, t, t)
\prod_{j=1}^{N} \left[
\frac{1}{\sigma \sqrt{2 \pi t}}
y_j^{-(N-1) \theta(t)/2-(N+1)/2}
\exp \left\{ - \frac{(\ln y_j)^2}{2 \sigma^2 t} \right\}
\right]
\nonumber\\
&& \qquad \times
\prod_{1 \leq j < k \leq N}
(y_k-y_j)(y_k^{\theta(t)}-y_j^{\theta(t)}).
\label{eqn:bio1}
\end{eqnarray}
If we change the variables as $y_j \to z_j$ by
\begin{equation}
\exp \left[ \frac{1}{2}
\Big\{ (N-1) \theta(t)+(N+1) \Big\} \sigma^2 t \right] y_j
=z_j, \quad 1 \leq j \leq N,
\label{eqn:change3}
\end{equation}
we obtain the probability density of the
distribution $\z=(z_1, z_2, \dots, z_N) 
\in \W_N^{\rm C} \equiv \{\z \in \R^N:
0 < z_1 < z_2 < \cdots < z_N\}$ 
(the Weyl chamber of type C$_N$), which is 
given by
\begin{equation}
P_N(t,\z)=C_N(a, \sigma, t)
\prod_{j=1}^N w(z_j)
\prod_{1 \leq j < k \leq N}
\Big\{(z_k-z_j) (z_k^{\theta(t)}-z_j^{\theta(t)}) \Big\},
\label{eqn:bio2}
\end{equation}
where
\begin{eqnarray}
&& C_N(a, \sigma, t)
=
\frac{1}{\prod_{n=1}^{N-1} (e^{n a \sigma}-1)^{N-n}}
\nonumber\\
&& \times
\exp \left[ - \frac{N}{12} \left\{
(N+1)(2N+1) \sigma^2 t 
+2(N^2-1) a \sigma
+(N-1)(2N-1) \frac{a^2}{t} \right\} \right],
\label{eqn:cN2}
\end{eqnarray}
and
\begin{equation}
w(z)=\frac{\beta}{\sqrt{\pi}}
e^{-\beta^2 (\log z)^2}
\quad \mbox{with} \quad
\beta=\frac{1}{\sigma \sqrt{2t}}.
\label{eqn:SW1}
\end{equation}
The probability density of the GUE (\ref{eqn:RM1})
is proportional to a square of the Vandermonde
determinant, while (\ref{eqn:bio2})
is proportional to the product of
$h_N(\z)$ and $h_N(\{z_j^{\theta}\})$.
The ensemble with the probability density
in this form (\ref{eqn:bio2}) is
called the biorthogonal ensemble and studied in
\cite{Mut95,Bor99}.
The special case with the weight function (\ref{eqn:SW1})
was studied in the name of
{\it the biorthogonal Stieltjes-Wigert matrix model}
for the Chern-Simons theory
by Dolivet and Tierz \cite{DT07,Tie10},
since (\ref{eqn:SW1}) is the weight function
for the Stieltjes-Wigert orthogonal polynomials
\cite{Sze81}.
In particular, when $t=t_0 \equiv a/\sigma$,
$\theta(t)$ becomes unity and the system is
reduced to the Stieltjes-Wigert matrix model
studied by Tierz \cite{Tie04}.
See Remark 3 in Sec.\ref{section:family}.
(Since the partition function $Z$ is given by 
a hermitian-matrix integral as Eq.(1.6) in \cite{DT07},
it is called the matrix model.
The partition function $Z$ is also written as 
the integral of weights of real eigenvalues as
Eq.(\ref{eqn:Z1}) below, then the matrix model
is identified with a statistical ensemble
of points on $\R$.)

\vskip 0.3cm
\noindent{\bf Remark 1} \,
In the theory of Chern-Simons matrix models
\cite{Mar05,dHT05,Tie04,DT07,Tie10}, the main quantity
to be calculated is the partition function, which 
will be given by
\begin{equation}
Z=\int_{\W_N^{\rm C}} 
\prod_{j=1}^N w(z_j)
\prod_{1 \leq j < k \leq N}
\Big\{(z_k-z_j) (z_k^{\theta(t)}-z_j^{\theta(t)}) \Big\} d \z.
\label{eqn:Z1}
\end{equation}
In the present setting, through the relations
(\ref{eqn:theta1}) and (\ref{eqn:SW1}), 
$Z$ is a function of $a, \sigma$ and $t$.
As a matter of course, in order to identify
(\ref{eqn:Z1}) with the partition function
studied in \cite{DT07},
we have to rewrite these parameters
by using the proper parameters
in the Chern-Simons theory,
but the partition function is essentially given by (\ref{eqn:Z1}).
Since $P_N(t, \z)$ given by (\ref{eqn:bio2})
is a probability density, it is normalized, and then
(\ref{eqn:Z1}) is equal to
the inverse of $C_N$ given by (\ref{eqn:cN2}).
Here we would like to put emphasis on the fact that,
since in the present work the ``matrix model" is
realized as a transform of the stochastic process
(the noncolliding Brownian motion with drift),
$C_N$ has been automatically obtained by 
the transformation from the normalization factor
$c_N$ given by (\ref{eqn:cN1}),
and $c_N$ is readily obtained by just putting
the proper conditions on the initial configuration
(\ref{eqn:equidistant}) and the drift vector (\ref{eqn:cond1})
as demonstrated in the proof of Proposition 1.
In \cite{DT07} the partition function was calculated
by using the biorthogonal Stieltjes-Wigert polynomials.
We do not need them to evaluate $C_N=1/Z$,
but we will also introduce them in the next section
in order to discuss the correlation functions
of the matrix model and our stochastic process.

\section{Time-Dependent Biorthogonal Stieltjes-Wigert Ensemble
as a Family of Determinantal Point Processes}
\label{chap:SW}
\subsection{Biorthonormal Stieltjes-Wigert Polynomials}

As functions of $q \in (0, 1)$, we consider
the $q$-extension of the Pochhammer symbol as
\begin{eqnarray}
(a;q)_0 &\equiv& 1,
\nonumber\\
(a;q)_n &\equiv& (1-a)(1-aq)(1-aq^2) \dots (1-aq^{n-1}),
\quad n \in \N, 
\nonumber
\end{eqnarray}
for $a \in \R$, and the $q$-binomial coefficients defined by
$$
\left[ \begin{array}{c} n \cr \ell \end{array} \right]_q
=
\frac{(q;q)_n}{(q;q)_{\ell} (q;q)_{n-\ell}},
\quad \ell=1,2, \dots, n-1,
$$
and 
$$
\left[ \begin{array}{c}
n \cr 0 \end{array} \right]_q
=
\left[ \begin{array}{c}
n \cr n \end{array} \right]_q =1.
$$

Let
\begin{equation}
w(z;q)=\frac{1}{\sqrt{2 \pi |\ln q|}}
\exp \left\{ - \frac{(\ln z)^2}{2 |\ln q|} \right\}.
\label{eqn:PSW3}
\end{equation}
The following two series of functions were introduced
in Appendix A.2 in \cite{DT07},
for $\theta \in (0, \infty), q \in (0,1)$,
\begin{eqnarray}
&& T_n(x; \theta, q)
= (-1)^n \frac{\sqrt{(q;q)_n} q^{(n \theta+1/2)/2}}
{(q^{\theta}; q^{\theta})_n}
\sum_{\ell=0}^{n} (-1)^{\ell}
\left[ \begin{array}{c} n \cr \ell \end{array} \right]_{q^{\theta}}
q^{\theta \ell(\ell(\theta+1)+1)/2} x^{\theta \ell},
\nonumber\\
&& R_n(x; \theta, q)
= (-1)^n \frac{q^{(n \theta+1/2)/2}}
{\sqrt{(q; q)_n}}
\sum_{\ell=0}^{n} (-1)^{\ell}
\left[ \begin{array}{c} n \cr \ell \end{array} \right]_{q^{\theta}}
q^{\ell(\ell(\theta+1)+(1-\theta)+1)/2} x^{\ell}, 
\label{eqn:TR1}
\end{eqnarray}
$n \in \N_0 \equiv \{0,1,2, \dots\}$,
where $T_n(x; \theta, q)$ is a polynomial of
$x^{\theta}$ of order $n$
and $R_n(x; \theta, q)$ is a polynomial
of $x$ of order $n$, $n \in \N_0$.

\noindent{\bf Proposition 2} \,
The polynomials 
$\{T_n(x; \theta, q), R_n(x; \theta, q)\}_{n \in \N_0}$
satisfy the following orthonormality relation
with respect to the weight function (\ref{eqn:PSW3}),
\begin{equation}
\int_0^{\infty} T_n(x;\theta,q) R_m(x; \theta, q)
w(x;q) dx =\delta_{n m},
\quad n, m \in \N_0.
\label{eqn:orthoTR}
\end{equation}
\vskip 0.3cm

\noindent{\bf Remark 2} \,
We can see that
\begin{equation}
\lim_{\theta \to 1} T_n(x; \theta, q)
=\lim_{\theta \to 1} R_n(x; \theta, q)
=p_n(x;q),
\quad n \in \N_0,
\label{eqn:SWlimit}
\end{equation}
where $p_n(x;q), n \in \N_0$ are
the Stieltjes-Wigert polynomials given by
\cite{Sze81}
\begin{equation}
p_n(x;q)=(-1)^n \frac{q^{(2n+1)/4}}{\sqrt{(q;q)_n}}
\sum_{\ell=0}^n 
\left[ \begin{array}{c} n \cr \ell \end{array} \right]_q
q^{\ell^2}(-q^{1/2} x)^{\ell},
\quad n \in \N_0.
\label{eqn:PSW4}
\end{equation}
(Note that the definition (\ref{eqn:PSW4})
is slightly different from that given in \cite{KS96}.)
Dolivet and Tierz \cite{DT07} derived
$\{T_n(x; \theta, q), R_n(x; \theta,q)\}_{n \in \N_0}$
by taking a limit $\alpha \to \infty$
of the $q$-Konhauser polynomials,
which has a parameter $\alpha$ in addition to
$\theta$ and $q$.
The weight function, with which the $q$-Konhauser polynomials
make orthogonality relations, is called the $q$-Laguerre measure
\cite{ASV83}.
We note that the $\alpha \to \infty$ limit of
the $q$-Laguerre measure is different from 
the present weight function $w(z;q)$ given by (\ref{eqn:PSW3}).
(It is related with the weight function
used in \cite{KS96} to define the Stieltjes-Wigert 
polynomials.) 
We give a proof of the orthonormality (\ref{eqn:orthoTR})
with respect to (\ref{eqn:PSW3}) below.
It implies that as well as the Stieltjes-Wigert
polynomials (\ref{eqn:PSW4}),
the biorthonormal polynomials 
$\{T_n(x; \theta, q), R_n(x; \theta,q)\}_{n \in \N_0}$
are indeterminate and there are many different
weight functions
(the Stieltjes moment problem, see \cite{Tie04,dHT05}).
\vskip 0.3cm
\noindent{\it Proof of Proposition 2} \,
It is enough to prove the following,
\begin{eqnarray}
\label{eqn:I1}
&& I_{n,m} \equiv
\int_0^{\infty} x^m T_n(x;\theta,q) w(x:q) dx
=t_{n}(\theta,q) \delta_{n m}, \quad 0 \leq m \leq n,
\\
\label{eqn:J1}
&& J_{n,m} \equiv
\int_0^{\infty} x^m R_n(x;\theta,q) w(x:q) dx
=r_{n}(\theta,q) \delta_{n m},
\quad 0 \leq m \leq n,
\end{eqnarray}
where
\begin{eqnarray}
&& t_{n}(\theta,q)
= \sqrt{(q;q)_n} q^{-(n+1/2)^2-(\theta-1)n^2/2},
\nonumber\\
&& r_{n}(\theta,q)
= \frac{(q^{\theta}; q^{\theta})_n}
{\sqrt{(q;q)_n}} 
q^{-(n\theta+1/2)^2+n^2 \theta(\theta-1)/2}.
\label{eqn:rt1}
\end{eqnarray}
It is easy to confirm 
\begin{equation}
\int_0^{\infty} x^{n} w(x;q) dx=
q^{-(n+1)^2/2},
\quad n \in \N_0
\label{eqn:basic1}
\end{equation}
for the weight function (\ref{eqn:PSW3}).
Then by definition of the polynomials
(\ref{eqn:TR1}), 
\begin{eqnarray}
\label{eqn:I2}
&& I_{n,m}=d_n \sum_{\ell=0}^{n}
\frac{(q^{-n \theta}; q^{\theta})_{\ell}}
{(q^{\theta}; q^{\theta})_{\ell}}
q^{(n-m) \ell \theta}
q^{-(m+1)^2/2},
\\
\label{eqn:J2}
&& J_{n,m}=\overline{d}_n \sum_{\ell=0}^{n}
\frac{(q^{-n \theta}; q^{\theta})_{\ell}}
{(q^{\theta}; q^{\theta})_{\ell}}
q^{(n-m) \ell \theta}
q^{-(m \theta+1)^2/2},
\end{eqnarray}
with
\begin{equation}
d_n=(-1)^n \frac{\sqrt{(q;q)_n} q^{(n\theta+1/2)/2}}
{(q^{\theta};q^{\theta})_n},\quad
\overline{d}_n=(-1)^n
\frac{q^{(n\theta+1/2)/2}}{\sqrt{(q;q)_n}}.
\label{eqn:ddbar1}
\end{equation}
The $q$-derivative of order $n$ of a function $f(x)$ is defined
as \cite{ASV83}
\begin{equation}
D_q^n f(x)
=\frac{1}{(1-q)^n x^n}
\sum_{\ell=0}^n \frac{(q^{-n};q)_{\ell}}{(q;q)_{\ell}}
q^{\ell} f(q^{\ell} x).
\label{eqn:Dq1}
\end{equation}
Then
\begin{eqnarray}
D_{q^{-\theta}}^n f(x)
&=& \frac{1}{(1-q^{-\theta})^n x^n}
\sum_{\ell=0}^n 
\frac{(q^{n\theta};q^{-\theta})_{\ell}}
{(q^{-\theta}; q^{-\theta})_{\ell}}
q^{-\ell \theta} f(x q^{-\ell \theta})
\nonumber\\
&=& \frac{1}{(1-q^{-\theta})^n x^n}
\sum_{\ell=0}^n 
\frac{(q^{-n\theta};q^{\theta})_{\ell}}
{(q^{\theta}; q^{\theta})_{\ell}}
q^{n\ell \theta} f(x q^{-\ell \theta}).
\label{eqn:Dq2}
\end{eqnarray}
Let $f(x)=x^m q^{-(m+1)^2/2}$ in (\ref{eqn:Dq2})
and compare the result with (\ref{eqn:I2}).
We obtain
\begin{equation}
I_{n,m}=d_n (1-q^{-\theta})^n
q^{-(m+1)^2/2} 
\left. D_{q^{-\theta}}^n x^m \right|_{x=1}.
\label{eqn:I3}
\end{equation}
It implies that $I_{n,m}=0$ for $0 \leq m < n$.
For $n=m$, by direct calculation, we see
$I_{n,n}=t_{n}(\theta,q)$.
Then (\ref{eqn:I1}) is proved.
For the proof of (\ref{eqn:J1}),
put $f(x)=x^m q^{-(m \theta+1)^2/2}$
in (\ref{eqn:Dq2}). We obtain 
\begin{equation}
J_{n,m}=\overline{d}_n (1-q^{-\theta})^n
q^{-(m \theta+1)^2/2}
\left. D_{q^{-\theta}}^n x^m \right|_{x=1}
\label{eqn:J3}
\end{equation}
and (\ref{eqn:J1}) is derived in the same way
as (\ref{eqn:I1}).
Then the proof is completed. \qed
\vskip 0.5cm

\subsection{Time-Dependent Stieltjes-Wigert Ensemble}

For $a, \sigma >0$, let 
\begin{equation}
\theta=\theta(t)=\frac{a}{\sigma t}, \quad
q=q(t)=e^{-\sigma^2 t}, \quad t \geq 0.
\label{eqn:thetaqA1}
\end{equation}
Then, by the equality between the Vandermonde
determinant and the product of differences
(\ref{eqn:Vand1}) and by the multi-linearity of
determinant, 
the probability density (\ref{eqn:bio2}) of $\z$
is written as follows,
\begin{equation}
P_N(t, \z)
=
\prod_{j=1}^N w(z_j; q(t))
\det_{1 \leq j, k \leq N}
\Big[ T_{k-1}(z_j; \theta(t), q(t)) \Big]
\det_{1 \leq \ell, m \leq N}
\Big[ R_{m-1}(z_{\ell}; \theta(t), q(t)) \Big].
\label{eqn:tdSW1}
\end{equation}
It is a one-parameter family with parameter $t \geq 0$
of the biorthogonal Stieltjes-Wigert ensembles.
We call it the 
{\it time-dependent biorthogonal Stieltjes-Wigert ensemble}.

\subsection{One-Parameter Family of Determinantal Point Processes 
Parameterized by Time}
\label{section:family}

Let ${\rm C}_0(\R_+)$ be the set of all continuous
real-valued functions with compact support
on $\R_+$.
For $f \in {\rm C}_0(\R_+)$ and $\kappa \in \R$,
at each fixed time $t \geq 0$, 
the generating function of correlation
functions is given by the following
Laplace transform of $P_N(t, \z)$,
\begin{eqnarray}
\cG_N(t, \chi)
&=& \frac{1}{N!}\int_{\R_+^N} e^{\kappa \sum_{j=1}^N f(z_j)}
P_N(t, \z) d \z
\nonumber\\
&=& \frac{1}{N!}\int_{\R_+^N} 
\prod_{j=1}^N (1-\chi(z_j)) P_N(t, \z) d \z,
\label{eqn:gN0}
\end{eqnarray}
where we put
\begin{equation}
\chi(z)=1-e^{\kappa f(z)}.
\label{eqn:chi1}
\end{equation}
If we write $\z_{N'}=(z_1, z_2, \dots, z_{N'})$, $N' \leq N$, 
the binomial expansion of the integrand of (\ref{eqn:gN0})
gives 
\begin{equation}
\cG_N(t, \chi)
= 1+\sum_{N'=1}^N (-1)^{N'} \frac{1}{N'!}
\int_{\R_+^{N'}}
\rho_N(t, \z_{N'})
\prod_{k=1}^{N'} \Big\{\chi(z_k) dz_k \Big\},
\label{eqn:gN1}
\end{equation}
where $\rho_N(t, \z_{N'})$ 
is the $N'$-point correlation function at time $t \geq 0$,
\begin{equation}
\rho_N(t, \z_{N'})
=\frac{1}{(N-N')!} \int_{\R^{N-N'}} P_N(t, \z) \prod_{j=N'+1}^N dz_j,
\quad 1 \leq N' \leq N.
\label{eqn:rho1}
\end{equation}

By the determinantal expression (\ref{eqn:tdSW1})
of the single-time probability density $P_N(t, \z)$,
(\ref{eqn:gN0}) is written as
\begin{eqnarray}
\cG_N(t, \chi)
&=& \frac{1}{N!} \int_{\R_+^N} 
\det_{1 \leq j, k \leq N}
\Big[ T_{k-1}(z_j; \theta(t), q(t)) w(z_j; q(t)) (1-\chi(z_j)) \Big]
\nonumber\\
&& \quad \times
\det_{1 \leq \ell, m \leq N}
\Big[R_{m-1}(z_{\ell}; \theta(t),q(t)) \Big] d \z,
\label{eqn:gN2}
\end{eqnarray}
where multi-linearity of determinants is used.
For square integrable functions
$g_{j}, \bar{g}_{j}, 1 \leq j \leq N$
on $\Lambda \subset \R$, 
the identity
\begin{equation}
\frac{1}{N!}
\int_{\Lambda^{N}} 
\det_{1 \leq j, k \leq N} \Big[ g_{j}(x_{k}) \Big]
\det_{1 \leq \ell, m \leq N} \Big[\bar{g}_{\ell}(x_{m}) \Big]
\, d \x
= \det_{1 \leq j, k \leq N} \left[
\int_{\Lambda} g_{j}(x) \bar{g}_{k}(x) dx \right],
\label{eqn:Andreief}
\end{equation}
is proved, which is called 
the Andr\'eief identity.
Then, if we set
\begin{eqnarray}
&& B_j(z)=T_{j-1}(z; \theta(t), q(t))
\sqrt{w(z; q(t))} \chi(z),
\nonumber\\
&& C_j(z)=R_{j-1}(z; \theta(t), q(t))
\sqrt{w(z; q(t))}, \quad j \in \N_0, 
\label{eqn:BC1}
\end{eqnarray}
we have the determinantal expression
\begin{equation}
\cG_N(t, \chi)
=\det_{1 \leq j,k \leq N}
\left[ \delta_{j k} 
- \int_{0}^{\infty} B_j(z) C_k(z) dz \right],
\label{eqn:gN3}
\end{equation}
where we have used the orthonormality
(\ref{eqn:orthoTR}) proved in Proposition 2.

Then, by Fredholm's expansion-formula of determinant
and by cyclic property and multi-linearity of determinants,
we have the equalities
\begin{eqnarray}
\cG_N(t, \chi) 
&=& 1+\sum_{N'=1}^N (-1)^{N'} \frac{1}{N'!}
\int_{\R_+^{N'}}
\det_{1 \leq j, k \leq N'}
\left[ \sum_{\ell=1}^N C_{\ell}(z_j) B_{\ell}(z_k) \right]
d \z_{N'}
\nonumber\\
&=& 1+\sum_{N'=1}^N (-1)^{N'} \frac{1}{N'!}
\int_{\R_+^{N'}} 
\det_{1 \leq j, k \leq N'} [K_N(t; z_j, z_k)]
\prod_{k=1}^{N'} \Big\{ \chi(z_k) dz_k \Big\}
\label{eqn:Fred1}
\end{eqnarray}
with the integral kernel for $(x,y) \in \R_+^2$ with $t \geq 0$
\begin{equation}
K_N(t; x,y)
= \sqrt{w(x;q(t)) w(y;q(t))}
\sum_{j=0}^{N-1} R_j(x; \theta(t),q(t))
T_j(y; \theta(t), q(t)).
\label{eqn:tdK1}
\end{equation}
The rhs of (\ref{eqn:Fred1}) is the definition of the 
Fredholm determinant, which is denoted by
\begin{equation}
\cG_N(t, \chi)
=\Det_{x,y \in \R_+}
\Big[ \delta(x-y)-K_N(t; x,y) \chi(y) \Big].
\label{eqn:Fred2}
\end{equation}
Comparing (\ref{eqn:Fred1}) with (\ref{eqn:gN1}),
we can conclude that, 
at each fixed time $t \geq 0$, for any $1 \leq N' \leq N$, 
the $N'$-point correlation function is given by
the determinant in the form
\begin{equation}
\rho_N(t, \z_{N'})
=\det_{1 \leq j, k \leq N'}
\Big[ K_N(t; z_j, z_k) \Big],
\quad 1 \leq N' \leq N, \quad t \geq 0.
\label{eqn:tddet1}
\end{equation}
In particular, the particle density
is given by the one-point correlation function
as
\begin{eqnarray}
\rho_N(t; z) 
&=& K_N(t, z,z), \quad z \geq 0, \quad t \geq 0.
\label{eqn:tddensity1}
\end{eqnarray}
The ensemble of points such that
any correlation function is given by
a determinant (and thus the generating
function of correlation functions is
given by a Fredholm determinant)
is called the {\it determinantal}
(or {\it fermion}) {\it point process} \cite{Sos00,ST03}.
The integral kernel $K_N$ is called the
{\it correlation kernel}.

\vskip 0.3cm
\noindent{\bf Remark 3} \,
For each $a > 0, \sigma >0$, there is a special time
\begin{equation}
t_0=\frac{a}{\sigma},
\label{eqn:t0_1}
\end{equation}
at which
\begin{equation}
\theta(t_0)=1, \quad
\sigma^2 t_0=a \sigma=\frac{a^2}{t_0}, \quad
q(t_0)=e^{-a \sigma} \equiv q_0.
\label{eqn:t0_2}
\end{equation}
At $t=t_0$, the probability density (\ref{eqn:tdSW1}) becomes
\begin{equation}
P_N(t_0; \z) =\widehat{c}_N(q_0)
\prod_{j=1}^N w(z_j;q_0)
\prod_{1 \leq j<k \leq N} (z_k-z_j)^2
\equiv P_N^{q_0}(\z)
\label{eqn:t0_3}
\end{equation}
with 
\begin{equation}
\widehat{c}_N(q_0)=\frac{q_0^{N(4N^2-1)/6}}
{\prod_{j=1}^{N-1} (q_0;q_0)_j}.
\label{eqn:t0_4}
\end{equation}
Then the system is reduced to the Stieltjes-Wigert
ensemble studied by \cite{Tie04},
which is also a determinantal point process with
the correlation kernel
\begin{equation}
K_N^{q_0}(x,y)=\sqrt{w(x;q_0) w(y;q_0)} 
\sum_{j=0}^{N-1} p_j(x;q_0) p_j(y;q_0),
\quad (x,y) \in \R_+^2,
\label{eqn:KN_t0}
\end{equation}
as derived from (\ref{eqn:tdK1}) by (\ref{eqn:SWlimit}).
For the Stieltjes-Wigert polynomial
$p_n$ given by (\ref{eqn:PSW4}), 
the three-term recurrence relation is given by
\begin{equation}
p_n(x;q)=\frac{q^{2n}x-q^{1/2}(1+q-q^n)}{\sqrt{1-q^n}}
p_{n-1}(x;q)-q^2 \frac{\sqrt{1-q^{n-1}}}{\sqrt{1-q^n}} p_{n-2}(x;q),
\quad n=2,3, \dots.
\label{eqn:pnrec}
\end{equation}
We can obtain the
Christoffel-Darboux formula for them,
\begin{eqnarray}
&&
\sum_{n=0}^{N-1} p_n(x;q) p_n(y;q)
=\frac{\sqrt{1-q^N}}{q^{2N}}
\frac{p_N(x;q)p_{N-1}(y;q)
-p_{N-1}(x;q) p_{N}(y;q)}{x-y},
\nonumber\\
&& \hskip 8cm \mbox{for $x \not= y$},
\nonumber\\
&&
\sum_{n=0}^{N-1} p_n(x)^2
=\frac{\sqrt{1-q^N}}{q^{2N}}
\Big\{ p_N'(x;q) p_{N-1}(x;q)
-p_{N-1}'(x;q) p_N(x;q) \Big\}.
\label{eqn:CD2}
\end{eqnarray}
Then the correlation kernel (\ref{eqn:KN_t0})
is rewritten as
\begin{eqnarray}
&&
K_N^{q_0}(x,y)
=\frac{\sqrt{1-q_0^N}}{q_0^{2N}} 
\sqrt{w(x;q_0)w(y;q_0)}
\frac{p_N(x;q_0)p_{N-1}(y;q_0)
-p_{N-1}(x;q_0) p_{N}(y;q_0)}{x-y},
\nonumber\\
\label{eqn:KCD1}
&& \hskip 8cm \mbox{for $x \not= y$},
\\
\label{eqn:KCD2}
&&
K_N^{q_0}(x,x)
=\frac{\sqrt{1-q_0^N}}{q_0^{2N}} w(x;q_0)
\Big\{ p_N'(x;q_0) p_{N-1}(x;q_0)
-p_{N-1}'(x;q_0) p_N(x;q_0) \Big\}.
\end{eqnarray}

\vskip 0.3cm
\noindent{\bf Remark 4} \,
Consider the system again
at the special time $t=t_0$, but here we write $q_0=q$
for simplicity.
Let 
\begin{equation}
z_j=q^{-3/2} \widetilde{z}_j \sqrt{2(1-q)} +q^{-1/2}
\quad \Longleftrightarrow \quad
\widetilde{z}_j=\frac{q^{3/2}z_j-q}{\sqrt{2(1-q)}},
\quad 1 \leq j \leq N.
\label{eqn:q1a1}
\end{equation}
Then we can show that
\begin{equation}
\lim_{q \to 1}
P_N^q(\z) \prod_{j=1}^N dz_j
=P_N^1(\widetilde{\z}) \prod_{j=1}^{N} d \widetilde{z}_j
\label{eqn:q1a2}
\end{equation}
with
\begin{equation}
P_N^1(\widetilde{\z})
=\frac{(1/2)^{-N^2/2}}
{(2 \pi)^{N/2} \prod_{j=1}^{N} \Gamma(j)}
\prod_{j=1}^N e^{-\widetilde{z}_j^2}
\prod_{1 \leq j < k \leq N}
(\widetilde{z}_k-\widetilde{z}_j)^2,
\quad \widetilde{\z} \in \W_N.
\label{eqn:q1a3}
\end{equation}
By (\ref{eqn:t0_2}), $q=q_0=e^{-a \sigma} \to 1$
as $a \sigma \to 0$.
Note that (\ref{eqn:q1a3}) is equal to the
special case with the variance $\sigma^2=1/2$
of the probability density of eigenvalue distribution
of the GUE,
\begin{equation}
P_N^{\rm GUE}(\vlambda)
=\frac{\sigma^{-N^2}}
{(2 \pi)^{N/2} \prod_{j=1}^{N} \Gamma(j)}
\prod_{j=1}^N e^{-\lambda_j^2/2 \sigma^2}
\prod_{1 \leq j < k \leq N}
(\lambda_k-\lambda_j)^2,
\quad \vlambda \in \W_N,
\label{eqn:GUE}
\end{equation}
(see (\ref{eqn:RM1}) in Sec.\ref{chap:introduction}.)
Define
\begin{eqnarray}
&&\widetilde{K}_N^q(\widetilde{x}, \widetilde{y})
\nonumber\\
&&
=K_N^q \Big(q^{-3/2} \widetilde{x} \sqrt{2(1-q)} +q^{-1/2},
q^{-3/2} \widetilde{y} \sqrt{2(1-q)}+q^{-1/2} \Big)
q^{-3/2} \sqrt{2(1-q)},
\label{eqn:tildeK1}
\end{eqnarray}
where $K_N^q(\cdot, \cdot)$ is given by (\ref{eqn:KN_t0})
with $q_0=q$.
The following asymptotics is established \cite{KS96}
\begin{equation}
\lim_{q \to 1}
\frac{\sqrt{(q;q)_n}}{\{(1-q)/2\}^{n/2}}
q^{-n/2-1/4} 
p_n \Big( q^{-3/2} \sqrt{2(1-q)} \, x+ q^{-1/2} \Big)
=H_n(x), \quad n \in \N_0,
\label{eqn:q1limit1}
\end{equation}
where $H_n(x), n \in \N_0$ are the Hermite polynomials,
$$
H_n(x) = n ! \sum_{k=0}^{[n/2]}
(-1)^k \frac{(2x)^{n-2k}}{k ! (n-2k)!}.
$$
Then, as expected, we can prove the following convergence,
\begin{equation}
\lim_{q \to 1}
\widetilde{K}_N^q(x, y) =\sum_{j=0}^{N-1} \varphi_j(x) \varphi_j(y),
\label{eqn:Hermite}
\end{equation}
where $\varphi_j(x)$ is the Hermite function, 
$
\varphi_j(x)=e^{-x^2/2}H_j(x)/\sqrt{2^j j! \sqrt{\pi}},
j \in \N_0$.

Combining the results in Remarks 3 and 4, we can say that
in the double limit $\theta \to 1$ and $q \to 1$,
the biorthogonal Stieltjes-Wigert ensemble 
is reduced to the GUE ensemble,
which is a determinantal point process with 
the Hermite kernel (\ref{eqn:Hermite})
under an appropriate scaling of variables
(\ref{eqn:q1a1}).

\section{Time Evolution of the Noncolliding Brownian Motion
with Drift}
\label{chap:evolution}
\subsection{Main Result}

By the sequence of transformations (\ref{eqn:change1})
and (\ref{eqn:change3}), each variable $z_j$ of the
time-dependent Stieltjes-Wigert ensemble (\ref{eqn:tdSW1})
is related with the variable
$x_j=x_j^{(1)}, t=t_1$ for the original noncolliding 
Brownian motion with drift given in Proposition 1 
in Sec.\ref{section:multi}.
That is, 
\begin{eqnarray}
z_j &=& \exp \left[ \sigma x_j
+\frac{1}{2}\{(N-1) \theta(t)+(N+1)\} \sigma^2 t \right]
\nonumber\\
&=& \exp \left[ \sigma x_j
+\frac{1}{2}(N-1) a \sigma+ \frac{1}{2} (N+1) \sigma^2 t \right],
\label{eqn:changeB1}
\end{eqnarray}
$1 \leq j \leq N$,
where the relation (\ref{eqn:theta1}) was used.
Since 
$dz_j=\sigma z_j dx_j, 1 \leq j \leq N$,
we arrive at the following main result
of the present paper, where we have used
the cyclic property of determinant.

\vskip 0.3cm
\noindent{\bf Theorem 3} \,
Consider the noncolliding Brownian motion with $N$ particles
started at the equidistant points (\ref{eqn:equidistant})
at time $t=0$ with the drift coefficients (\ref{eqn:cond1}).
At each time $t > 0$, the particle configuration
$\{X_j(t)\}_{j=1}^N$ is the determinantal point process
in the sense that any $N'$-point correlation function
of $\x_{N'}=(x_1, x_2, \dots, x_{N'})$,
$1 \leq N' \leq N$, is given by determinant
\begin{equation}
\widehat{\rho}_N(t, \x_{N'})
=\det_{1 \leq j, k \leq N'}
\Big[ \K_N(t; x_j, x_k) \Big].
\label{eqn:main1}
\end{equation}
Here the correlation kernel $\K_N$ is given by
\begin{eqnarray}
\K_N(t;x,y)
&=& K_N \Big(t; e^{\sigma x+(N-1) a \sigma/2
+(N+1) \sigma^2 t/2},
 e^{\sigma y+(N-1) a \sigma/2
+(N+1) \sigma^2 t/2} \Big)
\nonumber\\
&& \quad \times 
\sigma e^{\sigma(x+y)/2+(N-1) a \sigma/2+(N+1) \sigma^2 t/2},
\quad (x,y) \in \R^2, 
\label{eqn:bK1}
\end{eqnarray}
where $K_N(t; x,y)$ is given by
(\ref{eqn:tdK1}) with (\ref{eqn:theta1}).
\vskip 0.3cm

\subsection{Time-Evolution of Particle Density}
\label{section:time}

\begin{figure}
\includegraphics[width=0.6\linewidth]{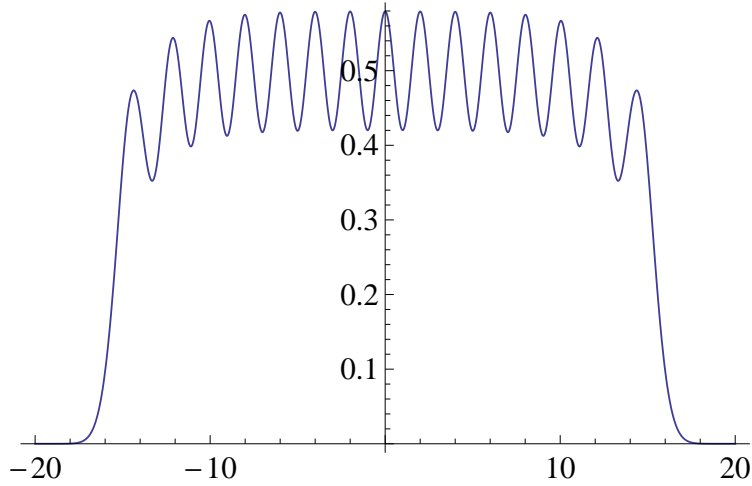}
\caption{The particle-density profile $\widehat{\rho}_N(t,x)$
of the noncolliding Brownian motion with drift at time $t=1$
for $N=15$, where $a=\sigma=1$.
The number of nodes is equal to $N=15$.
The oscillatory behavior represents the lattice structure
in one dimension caused by the present setting of
the equidistant initial configuration and well-ordered
drift coefficients.
}
\label{fig:Fig1}
\end{figure}

The particle density function for the $N$-particle process
is given by the one-point ($N'=1$) correlation function,
which is a `diagonal value' of the correlation kernel,
\begin{equation}
\widehat{\rho}_N(t, x)=\K_N(t; x,x),
\quad x \in \R, \quad t \geq 0.
\label{eqn:density1}
\end{equation}
Figure \ref{fig:Fig1} shows (\ref{eqn:density1})
for $N=15$ and $t=1, a=\sigma=1$,
{\it i.e.}, $\theta=1$ and $q=e^{-1} \simeq 0.37$.
The oscillatory behavior of density profile is
observed, which was already reported
in \cite{dHT05}.
In the present work, we put the equidistant initial
configuration (\ref{eqn:equidistant}) and the drift
coefficients which are regularly ordered as (\ref{eqn:cond1}),
and then the system has a lattice structure in one dimension
represented by this oscillatory behavior.

\begin{figure}
\includegraphics[width=0.6\linewidth]{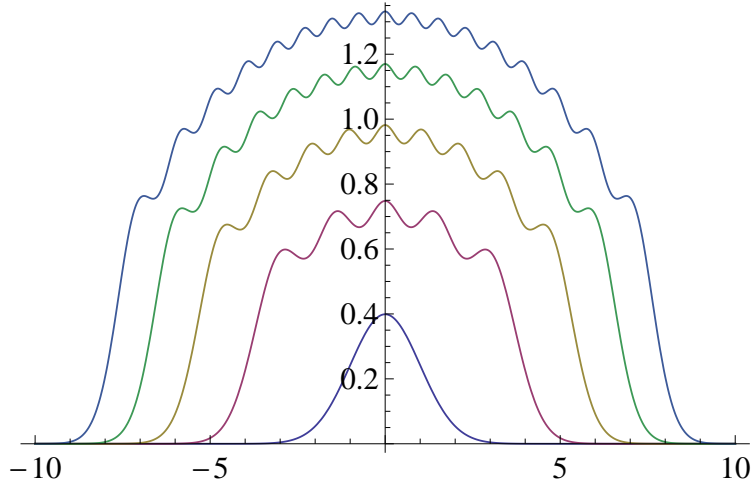}
\caption{The particle density functions
$\widetilde{\rho}_N(t,x)$ are plotted at time $t=1$
for $N=1,5,9,13,17$, from inner to outer,
for the noncolliding Brownian motion without drift
started at $\0$. As $N$ becomes large, the profile
becomes a semicircle with radius $\propto \sqrt{N}$.
}
\label{fig:Fig2}
\end{figure}

We compare the dependence on time $t$ and particle-number $N$
of density functions between
the present process, $\widehat{\rho}_N(t,x)$, and
the noncolliding Brownian motion without drift
started at $\0$, denoted by $\widetilde{\rho}_N(t,x)$.
As reviewed in Sec.5.1 in \cite{KT07},
the latter is given by
\begin{eqnarray}
\widetilde{\rho}_N(t,x)
&=& \frac{1}{\sqrt{2t}} \sum_{j=0}^{N-1}
\left\{ \varphi_j \left(\frac{x}{\sqrt{2t}}\right) \right\}^2
\nonumber\\
&=& \frac{1}{\sqrt{2t}}
\left[ N \left\{\varphi_N \left( \frac{x}{\sqrt{2t}} \right) \right\}^2
-\sqrt{N(N+1)} \varphi_{N-1} \left( \frac{x}{\sqrt{2t}} \right)
\varphi_{N+1} \left( \frac{x}{\sqrt{2t}} \right) \right],
\label{eqn:Wigner1}
\end{eqnarray}
$x \in \R, t >0$, where $\varphi_j$ is the
Hermite function, and it behaves asymptotically in $N \to \infty$ as
\begin{equation}
\widetilde{\rho}_N(t, x) \simeq
\left\{ \begin{array}{ll}
\displaystyle{\frac{1}{2t \pi}
\sqrt{(2 \sqrt{Nt})^2-x^2}},
& \quad \mbox{if $|x| \leq 2 \sqrt{N t}$}, \\
0, & \quad \mbox{if $|x| > 2 \sqrt{N t}$}. 
\end{array} \right.
\label{eqn:Wigner2}
\end{equation}
Figure \ref{fig:Fig2} shows (\ref{eqn:Wigner1}) at time $t=1$
for $N=1,5,9,13,17$ (from inner to outer).
We can observe oscillatory behavior also in this process
(the Dyson's Brownian motion model with $\beta=2$
started at $\0$), but the support of density function
of $N$ particles can expand only in the order $\sqrt{Nt}$.
Therefore, as shown in Fig.\ref{fig:Fig2}, 
the `wave length of oscillation' becomes
smaller as $\sqrt{t/N} \to 0$ in $N \to \infty$
for each time $0<t < \infty$ and we will have
Wigner's semicircle law (\ref{eqn:Wigner2})
for the particle-density profile 
asymptotically in $N \to \infty$.

\begin{figure}
\includegraphics[width=0.6\linewidth]{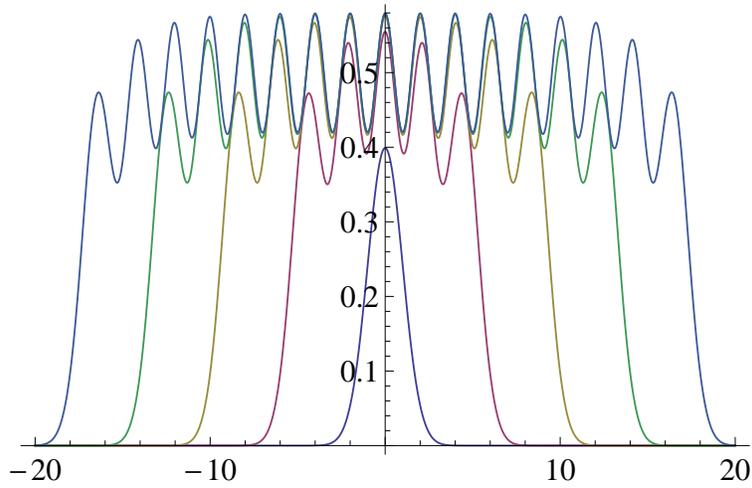}
\caption{The particle density functions
$\widehat{\rho}_N(t,x)$ are plotted at time $t=1$
for $N=1,5,9,13,17$, from inner to outer,
for the present noncolliding Brownian motion with drift
($a=\sigma=1$). 
As $N$ becomes large, the support of the profile
becomes wider proportionally to $N$.
Then the height of the profile is constant and
the lattice structure with $N$ nodes
is maintained.
}
\label{fig:Fig3}
\end{figure}

On the other hand, Fig.\ref{fig:Fig3} shows
the particle density functions (\ref{eqn:density1})
of the present noncolliding Brownian motion with drift
with $t=a=\sigma=1$ for $N=1,5,9,13,17$
(from inner to outer).

\begin{figure}
\includegraphics[width=0.4\linewidth]{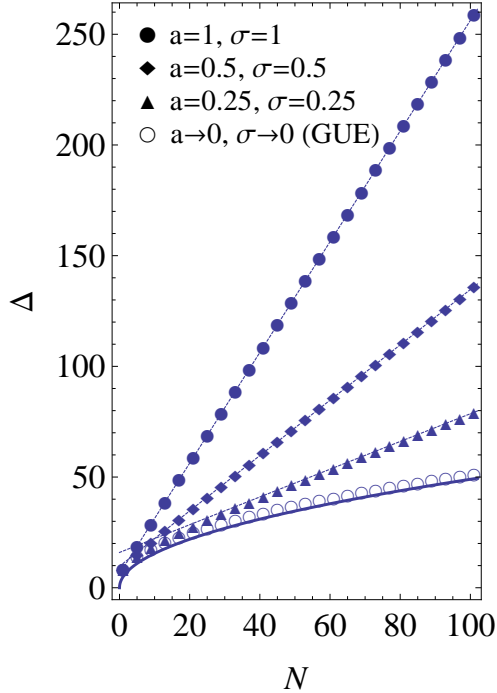}
\caption{The $N$-dependence of width of density profile 
$\Delta$ at time $t=1.5$
is shown for $a=\sigma=1$ (dotted by black circles),
$a=\sigma=0.5$ (by black squares),
$a=\sigma=0.25$ (by black triangles),
and for $a=\sigma=0$ (by white circles).
For large values of $N$, the dependence is well
described by a line 
$\Delta=c_1 N + c_2$ for the three cases
with $a=\sigma > 0$;
$(c_1, c_2)=(2.50, 6.29)$ for $a=\sigma=1$,
$(c_1, c_2)=(1.25, 9.70)$ for $a=\sigma=0.5$,
$(c_1, c_2)=(0.627, 15.9)$ for $a=\sigma=0.25$,
respectively.
The white circles for the noncolliding Brownian motion
without drift started at $\0$ approaches
to the result of Wigner's semicircle law for the GUE
described by a curve $\Delta=4 \sqrt{1.5 N} \simeq 4.9 \sqrt{N}$.
}
\label{fig:Fig4}
\end{figure}

As pointed out by de Haro and Tierz 
for the case with $\theta=1, q<1$ \cite{dHT05},
the width of support of profile increases 
proportionally to $N$, instead of $\sqrt{N}$.
In order to see $N$-dependence numerically,
here we define the width $\Delta(t)$ of density profile
at a given time $t>0$ as the length of interval of $x$
in which the value of $\widehat{\rho}_N$ (or $\widetilde{\rho}_N$) 
is greater than $\varepsilon=0.001$.
Figure \ref{fig:Fig4} shows the $N$-dependence of the
width at time $t=1.5$, $\Delta=\Delta(1.5)$, for the cases
$a=\sigma=1, 0.5, 0.25$
and for the case $a=\sigma=0$
(the noncolliding Brownian motion without drift
started at $\0$).
By (\ref{eqn:thetaqA1}), for the first three cases,
$\theta=1.5^{-1} \simeq 0.67$ in common
and $q \simeq 0.22, 0.69$, and 0.91, respectively.
In the cases with $a=\sigma >0$ ($q < 1$ and $\theta \not=1$ in general),
$\Delta$ increases linearly in $N$, while
it seems to approach to the curve
$\Delta = 4 \sqrt{1.5 N} \simeq 4.9 \sqrt{N}$ in the case
$a=\sigma=0$ as expected (Wigner's semicircle law).

Since we put drifts as (\ref{eqn:cond1}) to particles,
$t$-dependence of the support-width is $t$ (drifted), 
instead of $\sqrt{t}$ (diffusive).
In summary, we conjecture that, at each time $0 < t < \infty$,
\begin{equation}
\mbox{width of support of $\widehat{\rho}_N(t,x)
\propto Nt$ as $N \to \infty$.}
\label{eqn:conjecture1}
\end{equation}
Since $N$ particles exist on the support with
width $\propto N$, the average height of density profile
becomes independent of $N$ ($\propto 1/t$ in time $t$)
and the lattice structure (the oscillatory behavior
with $N$ nodes) will not become indistinct
even if $N$ is large.
Note that by noncolliding condition
the particles will show the fermionic (exclusive) behavior.
It is in contrast with Wigner's semicircle law
(\ref{eqn:Wigner2}), in which the height of profile
increases as $N$ becomes large; 
$\widetilde{\rho}_N(t,0) \propto \sqrt{N/t}$.

\section{Concluding Remarks}
\label{chap:conclusion}

In the present paper, we study the noncolliding Brownian motion
started at $N$ equidistant points (\ref{eqn:equidistant})
with a period $a$ for which drift coefficients
are chosen as (\ref{eqn:cond1}) with a scale $\sigma$ of values.
If we take the double limit $a \to 0$ and $\sigma \to 0$,
the process is reduced to the noncolliding Brownian motion
without drift started at $\0$.
It is well-known that, in this limit, 
the particle-position distribution is equivalent with
the eigenvalue distribution of random matrices
in GUE with variance $t$
and each time $t>0$ it is a determinantal point process
with the correlation kernel expressed by the Hermite
polynomials \cite{Meh04,For10}.
We have shown that, for any choice of positive values of
parameters $(a, \sigma)$,
the determinantal structure of particle distribution
is maintained, such that a pair of parameters
$(\theta, q)$ are determined at each time $t > 0$ by
\begin{eqnarray}
\label{eqn:qB1}
q &=& q(t)=e^{-\sigma^2 t},
\\
\label{eqn:thetaB1}
\theta &=& \theta(t)=\frac{a}{\sigma t},
\end{eqnarray}
and the correlation kernel $\K_N$ is expressed by
the $(\theta,q)$-extensions of the Hermite polynomials,
which are the biorthogonal Stieltjes-Wigert polynomials
introduced by Dolivet and Tierz \cite{DT07}.

As shown by (\ref{eqn:qB1}), if the system does not have
any drift term, $\sigma =0$, then $q \equiv 1$,
while if $\sigma >0$, then $q=1$ once at $t=0$
and $q$ decreases monotonically in time $t > 0$.
Introduction of drifts into the system is essential
for the present $q$-extension and the derivation
of value $q$ from 1 measures the time $t$.
In other words, the present stochastic process is
a system with a time-developing $q$-parameter (\ref{eqn:qB1}).
By (\ref{eqn:thetaB1}), we see that for each setting of
$(a, \sigma)$, there is a unique critical time $t_0=a/\sigma$
at which $\theta(t_0)=1$.

Generalization of Theorem 3 for time correlation functions
will be an interesting future problem.
The Eynard-Mehta-type correlation kernel
\cite{EM98,BR04,KT07,KT_cBM} shall be discussed.

Invalidity of Wigner's semicircle law for $q<1$
is an interesting phenomenon, which was first
observed by de Haro and Tierz in the case $\theta=1$ \cite{dHT05}
and is also reported for $\theta \not=1$ 
in Sec.\ref{section:time} in the present paper.
Asymptotic analysis of correlation kernel 
(\ref{eqn:bK1}) in $N \to \infty$ will be studied
so that the conjecture (\ref{eqn:conjecture1}) is proved.

It has been shown here that
the present noncolliding Brownian motion with drift
is exactly transformed to the biorthogonal
Stieltjes-Wigert matrix model studied by
Dolivet and Tierz \cite{DT07}.
We expect further connections in mathematics and physics
between nonequilibrium statistical mechanics
and the Chern-Simons theory in high-energy physics.

\begin{acknowledgments}
The present authors would like to thank
Gregory Schehr for useful comments on the present work
when he was invited to Chuo University in Tokyo (July 2012).
This work is supported in part by
the Grant-in-Aid for Scientific Research (C)
(No.21540397) of Japan Society for
the Promotion of Science.
\end{acknowledgments}

\appendix
\section{Drift Transform and the BBO Formula (\ref{eqn:BBO1})}
\label{chap:appendix_BBO}
First consider a one-dimensional standard 
Brownian motion without drift
started at the origin, $B(t), t\geq 0; B(0)=0$.
The probability density at time $t >0$ is given by
$u(t,x)=e^{-x^2/2t}/\sqrt{2 \pi t}$.
It solves the diffusion equation
$\partial u(t,x)/\partial t
=(1/2) \partial^2 u(t,x)/\partial x^2$
and we can see that
$u(0,x) \equiv \lim_{t \to 0} u(t, x)=\delta(x)$.
Let $\nu \in \R$ and consider the diffusion equation
with a drift term
\begin{equation}
\frac{\partial}{\partial t} u^{\nu}(t,x)
=\frac{1}{2} \frac{\partial^2}{\partial x^2} 
u^{\nu}(t, x)
+\nu \frac{\partial}{\partial x} u^{\nu}(t, x).
\label{eqn:diff2}
\end{equation}
A solution is given by
\begin{eqnarray}
u^{\nu}(t,x) &=& u(t, x+\nu t)
\nonumber\\
&=& \frac{1}{\sqrt{2 \pi t}}
\exp \left\{ -\frac{1}{2t} (x+\nu t)^2 \right\}.
\label{eqn:u2b}
\end{eqnarray}
The mean of the Gaussian distribution is shifted by
$-\nu t$, that is, $-\nu$ gives a drift velocity.
We note that (\ref{eqn:u2b}) is written as
$u^{\nu}(t, x)=e^{-\nu x-\nu^2t/2}u(t,x)$.
It is considered as follows (see, for instance, \cite{KS91}); 
$u^{\nu}(t,x)$ is
obtained from $u(t,x)$ by the {\it drift transform},
\begin{equation}
u(t,x) \, \to \, 
\exp \left( -\nu x -\frac{\nu^2}{2} t \right)
u(t,x).
\label{eqn:drift1}
\end{equation}
The transition probability density from $x$ to $y$
with time duration $t$ is then obtained by a shift of
spatial coordinate
\begin{eqnarray}
p^{\nu}(t, y|x)
&=& u^{\nu}(t, x-y)
\nonumber\\
&=& \exp \left\{ \nu(y-x)-\frac{\nu^2}{2} t \right\}
p(t,y|x),
\label{eqn:tpd1}
\end{eqnarray}
where $p(t,y|x)$ is given by (\ref{eqn:p0}).
Note that  
(\ref{eqn:tpd1}) satisfies the diffusion equation (\ref{eqn:diff2})
as a backward Kolmogorov equation.

Let $N=2,3,\dots$.
Consider the {\it vicious Brownian motion} defined as
\begin{eqnarray}
&& \mbox{$N$-particle system of Brownian motion killed when they collide} \,
\nonumber\\
&=& \mbox{$N$-dimensional Brownian motion in the Weyl chamber $\W_N$ with 
absorbing walls.}
\nonumber
\end{eqnarray}
The transition probability density 
from $\x \in \W_N$ to $\y \in \W_N$ with time duration $t \geq 0$
is given by the KM-LGV determinant $q_N(t, \y|\x)$, (\ref{eqn:KM1}).
It is a unique solution of the
differential equation
\begin{equation}
\frac{\partial}{\partial t} u(t, \x)
=\frac{1}{2} \Delta u(t, \x),
\quad \Delta \equiv \sum_{j=1}^N \frac{\partial^2}{\partial x_j^2},
\label{eqn:diffB1}
\end{equation}
satisfying the boundary condition
$u(t, \x)=0$ at $\x \in \partial \W_N$,
and the initial condition
$u(0, \x)=\delta(\x-\y)
=\prod_{j=1}^N \delta(x_j-y_j)$.

Now we consider the vicious Brownian motion problem with drift.
For $\vnu=(\nu_1, \dots, \nu_N) \in \R^N$, we want to solve
\begin{equation}
\frac{\partial}{\partial t} u(t, \x)
=\frac{1}{2} \Delta u(t, \x)
+ \vnu \cdot \nabla u(t, \x)
\label{eqn:diffC1}
\end{equation}
with the conditions
\begin{eqnarray}
&& u(t, \x) > 0, \quad \x \in \W_N, t >0, 
\nonumber\\
&& u(t, \x)=0 \quad \mbox{at $\x \in \partial \W_N, t >0$},
\nonumber\\
&& u(0, \x)=\delta(\x-\y).
\label{eqn:diffC3}
\end{eqnarray}
The solution is given by the drift transform
of the KM-LGV determinant (\ref{eqn:KM1}),
\begin{eqnarray}
q_N^{\vnu}(t, \y|\x)
&=& \exp \left\{ \vnu \cdot (\y-\x)
-\frac{|\vnu|^2}{2} t \right\}
q_N(t, \y|\x)
\nonumber\\
&=& \exp \left\{ \vnu \cdot (\y-\x)
-\frac{|\vnu|^2}{2} t \right\}
\det_{1 \leq j, k \leq N} \Big[
p(t, y_j|x_k) \Big].
\label{eqn:qNn1}
\end{eqnarray}

The important fact is that
$$
q_N^{\vnu}(t, \y|\x)
\not=
\det_{1 \leq j, k \leq N} \Big[
p^{\nu_j}(t, y_j|\x_k) \Big].
$$
That is, it is not the KM-LGV determinant
of the drift transform of $p$'s.
Actually we can see
\begin{eqnarray}
q_N^{\vnu}(t, \y|\x)
&=& e^{-\vnu \cdot \x}
\sum_{\sigma \in \cS_N} {\rm sgn}(\sigma)
\prod_{k=1}^N e^{\nu_k x_{\sigma(k)}}
\prod_{j=1}^N p(t,y_j-\nu_j t| x_{\sigma(j)})
\nonumber\\
&=& e^{-\vnu \cdot \x}
\sum_{\sigma \in \cS_N} {\rm sgn}(\sigma)
\prod_{k=1}^N e^{\nu_k x_{\sigma(k)}}
\prod_{j=1}^N p^{\nu_j}(t,y_j| x_{\sigma(j)}),
\label{eqn:qNn2}
\end{eqnarray}
where $\cS_N$ is the set of all permutations $\{\sigma\}$
of $N$ items.

The survival probability at time $t>0$ will be given by
\begin{eqnarray}
\cN_N^{\vnu}(t, \x)
&=& \int_{\W_N} q_N^{\vnu}(t, \y|\x) d \y
\nonumber\\
&=& e^{-\vnu \cdot \x}
\sum_{\sigma \in \cS_N} {\rm sgn}(\sigma)
\prod_{k=1}^N e^{\nu_k x_{\sigma(k)}}
\int_{\W_N} \prod_{j=1}^N p^{\nu_j}(t, y_j|x_{\sigma(j)}) d \y.
\label{eqn:NNn1}
\end{eqnarray}
It was claimed in \cite{BBO05} that
\begin{equation}
\mbox{if $\vnu \in \W_N$ and $\x \in \W_N$}
\label{eqn:condB1}
\end{equation}
then
\begin{equation}
\lim_{t \to \infty} \int_{\W_N} 
 \prod_{j=1}^N p^{\nu_j}(t, y_j|x_{\sigma(j)}) d \y
=1, \quad \forall \sigma \in \cS_N.
\label{eqn:lim1}
\end{equation}
That is, for any initial configuration $\x \in \R^N$,
if (\ref{eqn:condB1}) is satisfied,
\begin{equation}
\lim_{t \to \infty} \B^{\x, \vnu}(t) \in \W_N
\quad \mbox{with probab. 1.}
\label{eqn:limit2}
\end{equation}
It is a matter of course, since we put the drift vector
$\vnu \in \W_N$, the position vector of
Brownian motions should be in $\W_N$
if we wait for sufficiently long term,
independently of initial configuration.
Then we obtain \cite{BBO05}
\begin{eqnarray}
\lim_{t \to \infty} \cN_N^{\vnu}(t, \x)
&=& e^{-\vnu \cdot \x}
\sum_{\sigma \in \cS_N} {\rm sgn}(\sigma)
\prod_{k=1}^N e^{\nu_k x_{\sigma(k)}}
\nonumber\\
&=& e^{-\vnu \cdot \x}
\det_{1 \leq j, k \leq N} \Big[ e^{\nu_j x_k} \Big].
\label{eqn:BBO05}
\end{eqnarray}

\vskip 0.3cm
\noindent{\bf Remark 5} \,
The following argument is found in \cite{OCo12b}.
Let $S_N^{\vnu}(\x)$ be the survival probability
for the noncolliding BM with drift $\vnu \in \W_N$
started at $\x \in \W_N$.
Then, it is a stationary solution of the diffusion equation
\begin{equation}
0=\frac{1}{2} \Delta S_N^{\vnu}(\x) + \vnu \cdot \nabla S_N^{\vnu}(\x),
\label{eqn:diffS1}
\end{equation}
satisfying the conditions
\begin{eqnarray}
&& S_N^{\vnu}(\x) =0 \quad
\mbox{at $\x \in \partial \W_N$},
\nonumber\\
\label{eqn:S3}
&& \lim_{\x \to \infty} S_N^{\vnu}(\x)=1,
\end{eqnarray}
where
$$
\x \to \infty \quad
\Longleftrightarrow \quad
x_{j+1}-x_j \to \infty,\quad
1 \leq j \leq N-1.
$$
We can confirm that
$$
e^{-\vnu \cdot \x}
\det_{1 \leq j, k \leq N} \Big[ e^{\nu_j x_k} \Big]
$$
is the unique solution of this problem.
\vskip 0.3cm

The noncolliding Brownian motion with drift is defined as the
system of Brownian motions conditioned that 
they never collide with each other forever.
Then, the transition probability density is 
obtained by 
\begin{eqnarray}
p_N^{\vnu}(t,\y|\x)
&=& \lim_{T \to \infty}
\frac{\cN^{\vnu}(T-t, \y) q_N^{\vnu}(t,\y|\x)}{\cN^{\vnu}(T, \x)}
\nonumber\\
&=& 
\frac{e^{- \vnu \cdot \y} \displaystyle{
\det_{1 \leq j, k \leq N}
[e^{\nu_j y_k}]}}
{e^{- \vnu \cdot \x} \displaystyle{
\det_{1 \leq j, k \leq N}
[e^{\nu_j x_k}]}}
e^{\vnu \cdot(\y-\x)-|\vnu|^2 t/2}
q_N(t, \y|\x).
\nonumber
\end{eqnarray}
It is equal to (\ref{eqn:BBO1}).

\section{A Combinatorial Limit of O'Connell Process}
\label{chap:appendix_OCo}
Recently O'Connell introduced an interacting
diffusive particle system \cite{OCo12a}.
Let $\psi_{\vnu}^{(N)}(\x), \x \in \R^N, \vnu \in \C^N$
(the $N$-dimensional complex space) 
be the class-one Whittaker function,
whose Givental integral representation is
given by
\begin{eqnarray}
\psi_{\vnu}^{(N)}(\x)
&=&\int_{\Gamma_N(\x)} 
\exp \left[ 
\sum_{j=1}^{N} \nu_{j}
\left( \sum_{k=1}^{j} T_{j, k}
-\sum_{k=1}^{j-1} T_{j-1, k} \right)
\right. \nonumber\\
&& \quad \left.
- \sum_{j=1}^{N-1} \sum_{k=1}^j
\Big\{ e^{-(T_{j,k}-T_{j+1,k})}
+e^{-(T_{j+1, k+1}-T_{j,k})} \Big\} \right] d \T,
\label{eqn:WInt2}
\end{eqnarray}
where the integral is performed 
over the space of all real lower
triangular arrays with size $N$,
$\T=(T_{j,k}, 1 \leq k \leq j \leq N)$
conditioned $T_{N,k}=x_k, 1 \leq k \leq N$.
The transition probability density 
$P_N^{\vnu, \, a}$ of the O'Connell process
with a parameter $a>0$ \cite{Kat12a}
is a unique solution of the equation
\begin{equation}
\left[ \frac{\partial}{\partial t}
-\left( \frac{1}{2} \Delta
+\nabla \log \psi^{(N)}_{\vnu}(\x/a) \cdot \nabla \right) \right]
P_N^{\vnu, \, a}(t, \y|\x)=0
\label{eqn:Kol2}
\end{equation}
with the initial condition 
$P_N^{\vnu, \, a}(0, \y|\x)=\delta(\x-\y)$.
The solution is given by
\begin{equation}
P_N^{\vnu, \, a}(t, \y|\x)
=e^{-t|\vnu|^2/2 a^2}
\frac{\psi^{(N)}_{\vnu}(\y/a)}{\psi^{(N)}_{\vnu}(\x/a)}
Q_N^{a}(t, \y|\x)
\label{eqn:pNnuxi1}
\end{equation}
with
\begin{equation}
Q_N^{a}(t, \y|\x)
=\int_{\R^N} e^{-t|\k|^2/2}
\psi^{(N)}_{i a \k}(\x/a) \psi^{(N)}_{-i a \k}(\y/a)
s_N(a \k) d \k,
\label{eqn:QN1}
\end{equation}
where $i=\sqrt{-1}$ and 
$s_N(\vmu)$ is the density function of
the Sklyanin measure
\begin{equation}
s_N(\vmu)
=\frac{1}{(2 \pi)^N!}
\prod_{1 \leq j < \ell \leq N}
|\Gamma(i(\mu_{\ell}-\mu_j))|^{-2}.
\label{eqn:sN1}
\end{equation}

We can show that (\ref{eqn:Kol2}) is 
a geometric lifting with parameter $a >0$  of 
the diffusion equation (the backward Kolmogorov equation)
of the noncolliding Brownian motion with drift \cite{Kat11,Kat12a}.
Then the BBO formula (\ref{eqn:BBO1}) is regarded as 
a combinatorial limit of the transition probability density
of the O'Connell process in the sense that
\cite{Kat12b,Kat12c}
\begin{equation}
p_N^{\vnu}(t, \y|\x)
=\lim_{a \to 0}
P_N^{a \vnu, \, a}(t, \y|\x),
\quad
\x, \vnu \in \overline{\W}_N, \,
\y \in \W_N, \, t \in [0, \infty).
\label{eqn:pNnu3}
\end{equation}

We note that the above argument (the geometric lifting and
combinatorial limit) gives another proof
of the BBO formula (\ref{eqn:BBO1})
for the noncolliding Brownian motion with drift.

\section{Geometric Brownian Motion}
\label{chap:appendix_GBM}

For $\sigma \not=0, \mu \in \R$, and $x > 0$
consider the linear stochastic differential equation (SDE)
\begin{equation}
dX(t)= \sigma X(t) dB(t) + \mu X(t) dt, \quad
X(0)=x, \quad t \geq 0,
\label{eqn:SDE1}
\end{equation}
where $B(t), t \geq 0$ is a standard Brownian motion
started at 0.
The parameters $\sigma$ and $\mu$ are called
percentage volatility and percentage drift, respectively.
The process $X(t), t \geq 0$ is called 
a geometric (or exponential)
Brownian motion \cite{BS02}.
The backward Kolmogorov equation for the 
process $X(t)$ is given by
\begin{equation}
\frac{\partial}{\partial t} u(t,x)
=\frac{1}{2} \sigma^2 x^2 
\frac{\partial^2}{\partial x^2} u(t,x)
+ \mu x \frac{\partial}{\partial x} u(t,x).
\label{eqn:K2}
\end{equation}

Set
\begin{equation}
\widetilde{\nu}= \frac{1}{\sigma^2} \left(\mu-\frac{\sigma^2}{2} \right)
\quad \Longleftrightarrow \quad
\mu=\frac{2\widetilde{\nu}+1}{2} \sigma^2, 
\label{eqn:nu1}
\end{equation}
and put
\begin{equation}
p(t; x,y)=\frac{|\sigma|}{2 \sqrt{2 \pi t}}
(xy)^{-\widetilde{\nu}}
\exp \left(
-\frac{\sigma^2 \widetilde{\nu}^2 t}{2}
-\frac{(\ln y-\ln x)^2}{2 \sigma^2 t} \right).
\label{eqn:p1}
\end{equation}
It is easy to confirm that (\ref{eqn:p1})
satisfies the backward Kolmogorov equation (\ref{eqn:K2}).
Remark that $p(t; x,y)$ is symmetric with respect to
$x$ and $y$; 
$p(t; x,y)=p(t; y, x)$.

Let
\begin{equation}
m(dx)=\frac{2}{\sigma^2} x^{2\widetilde{\nu}-1} dx,
\label{eqn:m1}
\end{equation}
which is called the
speed measure of $X(t)$.
By the general theory of one-dimensional
diffusion processes \cite{BS02},
it is proved that for any $A \subset \R$ 
the probability
$
P_t(x,A)={\rm Prob}(X(t) \in A |
X(0)=x)
$
is given by
$
P_t(x, A)= \int_{A} p(t; x,y)
m(dy).
$
In other words, if we set
\begin{equation}
p(t, y|x)=p(t; x,y)
\frac{m(dy)}{dy},
\label{eqn:tp0}
\end{equation}
then $p(t, y|x)$ is the transition probability
density of the process from $x \in \R_+$
to $y \in \R_+$ during time $t>0$;
$
P_t(x,A)=\int_{A} p(t, y|x) dy.
$
From (\ref{eqn:p1}) and (\ref{eqn:m1}), we have
the following expressions,
\begin{eqnarray}
p(t, y|x) &=&
\frac{2}{\sigma^2} y^{2\widetilde{\nu}-1}
\frac{|\sigma|}{2 \sqrt{2 \pi t}}
(xy)^{-\widetilde{\nu}} 
\exp \left( - \frac{\sigma^2 \widetilde{\nu}^2 t}{2}
-\frac{(\ln y-\ln x)^2}{2 \sigma^2 t} \right)
\nonumber\\
&=& \frac{1}{y |\sigma| \sqrt{2 \pi t}}
\left( \frac{y}{x} \right)^{\widetilde{\nu}}
\exp \left( - \frac{\sigma^2 \widetilde{\nu}^2 t}{2}
-\frac{(\ln y-\ln x)^2}{2 \sigma^2 t} \right)
\nonumber\\
&=& \frac{1}{y |\sigma| \sqrt{2 \pi t}}
\exp \left\{
- \frac{1}{2 \sigma^2 t}
\Big(\ln(y/x)-\widetilde{\nu} \sigma^2 t \Big)^2 \right\}.
\label{eqn:tp1}
\end{eqnarray}

Let
\begin{eqnarray}
\sigma_t &=& |\sigma| \sqrt{t}, \nonumber\\
\mu_t(x) &=& \ln x + \widetilde{\nu} \sigma^2 t
= \ln x + \widetilde{\nu} \sigma_t^2.
\label{eqn:sigmat}
\end{eqnarray}
Then (\ref{eqn:tp1}) is written as
\begin{equation}
p(t,y|x) 
= \frac{1}{y \sigma_t \sqrt{2\pi}}
\exp \left\{ - 
\frac{(\ln y-\mu_t(x))^2}{2 \sigma_t^2} \right\}.
\label{eqn:LN1}
\end{equation}
It is nothing but the 
log-normal distribution with
parameters $\sigma_t$ and $\mu_t(x)$.

In the present paper we consider the case
\begin{equation}
\widetilde{\nu}=0 \quad \Longleftrightarrow \quad
\mu=\frac{\sigma^2}{2}.
\label{eqn:volatility1}
\end{equation}
That is, the SDE is given by
\begin{equation}
dX(t)=\sigma X(t) dB(t)
+\frac{\sigma^2}{2} X(t) dt, \quad
t \geq 0,
\label{eqn:SDE2}
\end{equation}
By a simple application of
It\^o's formula \cite{KS91}, we can confirm that
the solution of this SDE is given by 
(\ref{eqn:geoBM1}).

\section{Noncolliding Geometric Brownian Motion}
\label{chap:appendix_noncGBM}

For arbitrary $M \in \N, 0 < t_1 < \cdots < t_M < \infty$, 
the multitime joint probability density for the 
noncolliding Brownian motion without drift is given by
\cite{KT07,KT10}
\begin{eqnarray}
&& p_N(t_1, \x^{(1)}; \dots; t_M, \x^{(M)}|\x)
\nonumber\\
&& = \prod_{1 \leq j < k \leq N}
(x^{(M)}_k-x^{(M)}_j) 
\prod_{m=1}^{M-1} q_N(t_{m+1}-t_m, \x^{(m+1)}|\x^{(m)})
\frac{q_N(t_1, \x^{(1)}|\x)}
{\prod_{1 \leq j < k \leq N}(x_k-x_j)}
\label{eqn:pxi1}
\end{eqnarray}
for any fixed initial configuration
$\x=(x_1, \dots, x_N) \in \overline{\W}_N$.

By the transformation (\ref{eqn:geoBM1})
from $B(t) \to X(t), t \geq 0$,
the $N$-particle system of geometric Brownian motions
with percentage volatility $\sigma >0$
conditioned never to collide with each other,
which we call the {\it noncolliding geometric 
Brownian motion}, should have the
multitime joint probability density in the form
\begin{eqnarray}
&& p_N(t_1, \y^{(1)}; \dots; t_M, \y^{(M)}|\y)
= \prod_{1 \leq j < k \leq N}
(\ln y^{(M)}_k- \ln y^{(M)}_j) 
\nonumber\\
&& \qquad \qquad \times
\prod_{m=1}^{M-1} q_N^{\rm geo}(t_{m+1}-t_m, \y^{(m+1)}|\y^{(m)})
\frac{q_N^{\rm geo}(t_1, \y^{(1)}|\y)}
{\prod_{1 \leq j < k \leq N}
(\ln y_k- \ln y_j)}.
\label{eqn:pxiG1}
\end{eqnarray}
It does not seem to be possible to derive
(\ref{eqn:mpC1}) from this general formula
by just choosing an initial configuration $\y$.



\end{document}